# A 160°×160° Dynamic Holographic Meta-Projector

*Feng-Jun Li[1,#], Ruixing Xia[2,#], Qianmei Deng[1,#], Yuze Lu[2,#], Xiangping Li[1], Fangwen Sun[2], Dong Zhao[2,\*], Zi-Lan Deng[1,\*], Kun Huang[2,3,\*]*

[1]Guangdong Provincial Key Laboratory of Optical Fiber Sensing and Communications, Institute of Photonics Technology, College of Physics & Optoelectronic Engineering, Jinan University, Guangzhou, 510632, China

[2]Department of Optics and Optical Engineering, University of Science and Technology of China, Hefei 230026, China

[3]National Key Laboratory of Opto-Electronic Information Acquisition and Protection Technology, Anhui University, Hefei, 230601, China

[#] F. L., R. X., Q. D. and Y. L. contributed equally to this work.

[*] Corresponding authors: D. Z. (zd10375@ustc.edu.cn), Z.-L. D. (zilandeng@jnu.edu.cn), or K. H. (huangk17@ustc.edu.cn)

## Abstract

**Holography can reconstruct immersive light fields for virtual and augmented reality by modulating optical wavefront. Due to huge pixel sizes, current spatial light modulators (SLMs) have small field-of-view (FOV) for holographic displays. Despite various methods for etendue expansion, the largest full-screen FOV for dynamic holography is only ~70°×70°, which remains insufficient for large-scale, high-resolution, three-dimensional displays. Here, we report a pixel-interpolation-assisted holographic meta-projector that substantially expands the FOV by integrating multiple subwavelength metasurface pixels within each microscale pixel of a traditional SLM. Leveraging large-angle diffraction of the metasurface and implementing k-space distortion correction for ultra-wide angles, we experimentally demonstrate dynamic holographic image reconstruction with a FOV of ~160°×160° — equivalent to a system numerical aperture of 0.985—at a high framerate of 60 Hz, surpassing the temporal resolution threshold of human vision. This system represents the state-of-the-art near-full-screen holographic dynamic display, thereby opening the door to high-dynamic-range and large-FOV holographic displays.**

## Introduction

Holography, originally conceived for microscopy, enables object reconstruction through coherent illumination[1]. While traditional holograms recorded physical interference patterns and were limited to static displays[2], the advent of computer-generated holography[3-5] has enabled dynamic light-field control using spatial light modulators (SLMs)[6,7] or digital micromirror devices[8]. A fundamental constraint, however, persists: the micron-scale pixel pitch of these modulators inherently restricts the diffraction angle[9], leading to a narrow FOV in modern holographic systems[10-12]. This limitation arises from a physical trade-off—further pixel miniaturization is impeded by inter-pixel crosstalk, which becomes pronounced as pixel dimensions approach the transitional scale of liquid crystal molecules between adjacent units[13,14].

Multiple strategies have been explored to overcome this FOV barrier, including the use of random scattering medium[15], photon sieves[16,17], optical gratings[18-21], artificial phase masks[22,23] and high-NA meatlens[24]. While these approaches have advanced dynamic holographic display capabilities, their achievable FOVs remain modest—typically spanning from about 6°×6° [23] to 70°×70° [25]. These advances remain inadequate for applications requiring larger viewing angles to faithfully reconstruct high-resolution, three-dimensional objects.

Metasurfaces employing subwavelength-pixel architectures inherently offer large-FOV capabilities for diverse optical systems[26-37]. Yet, their practical utility in dynamic applications is hampered by limited reconfigurability—a consequence of the inherent difficulty in modulating optical properties, particularly the refractive index, within solid-state nanostructures[38] (Fig. 1a). Several approaches have been proposed to overcome this limitation, including stretchable substrates for pixel scaling[39], tunable surrounding media[40,41] or substrates[42,43], adjustable resonant cavities[44,45], phase-change material[46,47] and controllable chemical reactions[48]. Nevertheless, these approaches still face restrictions: either they permit only a limited number of discrete states[39-45], or they suffer from impractically slow refresh rates (<1 Hz [46,47]), rendering them unsuitable for real-time holographic display. Alternatively, switchable illumination on $N$ static meta-holograms could reconstruct $2^N$ combined patterns for dynamic display[49,50]. However, over vast majority (>99%) of these patterns convey no meaningful information. While this limitation can be partially mitigated by illuminating a metasurface with a phase-encoded beam—effectively matching the incident phase pitch to that of the metasurface, thereby enabling arbitrary pattern generation[51,52]. This configuration causes the metasurface to behave as a conventional diffractive element, ultimately resulting in a narrow FOV [51]. Thus, the realization of dynamic meta-holography that combines a wide FOV with the ability to reconstruct arbitrary images remains an unresolved challenge at the intersection of conventional optics and metasurface research.

Here we introduce a pixel-interpolation-assisted dynamic holographic meta-projector that overcomes previous field-of-view limitations by integrating multiple discrete subwavelength metasurface pixels within each microscale pixel of a spatial light modulator (SLM). This design merges the capacity of interpolated metasurfaces to govern high spatial frequencies with the high-dynamic-range phase modulation offered by SLMs. Following pixel interpolation, the resulting composite phase profile simultaneously exhibits wide diffraction angles and high phase reconfigurability, enabling the reconstruction of holographic video with a field of view up to 160°×160°—closely approaching the theoretical full-screen limit—at a refresh rate of 60 Hz, substantially beyond the temporal resolution of human vision. This methodology establishes a viable route toward high-dynamic-range, wide-FOV holographic displays.

**Principle of meta-pixel interpolation**

Ideal reconfigurable meta-holography realizes dynamic control of light through active metasurfaces (Fig. 1a), which, however, are not achievable nowadays due to poor tunability of nanostructures. Instead, our pixel-interpolation approach achieves dynamic control via a static metasurface combined with spatially structured illumination (Fig. 1b). In this scheme, dynamic holographic reconstruction is enabled by a phase-modulated incident beam generated by a conventional spatial light modulator (SLM). Although the SLM imposes a phase profile with micron-scale pixelation, this pattern is projected onto the underlying static metasurface, where each SLM pixel corresponds to a cluster of distinct subwavelength-phase shifting elements (Fig. 1c). The resulting interpolated phase combines dynamic programmability with effective subwavelength spatial resolution. This in turn enables the generation of high spatial frequencies in the diffracted wavefront, supporting large diffraction angles essential for FOV operation. Consequently, our pixel-interpolation framework not only facilitates the demonstration of large-FOV holographic video but also establishes the physical basis for wide-angle dynamic meta-holography.

Since the dynamic phase comes from only the large-pixel SLM, the interpolated phase is quasi-dynamic, offering a lower degree of tunability than that of an ideal dynamic metasurface hologram. To compensate for this limitation, the metasurface phase profile is pre-optimized to meet specific operational requirements. In this work, for instance, the metasurface is engineered with an expanded phase profile to achieve a record full-screen FOV of 160°×160° (Fig. 1d), as will be experimentally validated in later sections. Consequently, when a sufficiently high-resolution SLM is employed—such as the 2000×2000 pixel device used here—the dynamic SLM phase alone can be tasked with reconstructing arbitrary holographic images.

A pixel-compression process is required prior to pixel interpolation, in which the SLM phase pattern is optically projected onto the metasurface via an imaging system (Fig. 2a). This step is essential for reflective-type SLMs, such as those based on liquid-crystal-on-silicon technology. It also enables effectively zero spatial separation between the SLM and metasurface phase planes. Furthermore, pixel compression reduces the effective SLM pixel size by a scaling factor determined by the imaging optics. Notably, since the entire active area is scaled proportionally, the space-bandwidth product (SBP) of the SLM remains unchanged. This proportional scaling also allows a corresponding reduction in the metasurface dimensions, simplifying its fabrication. We note that pixel compression would be unnecessary if the metasurface could be fabricated at the same scale as the SLM. Figure 2b summarizes the distinct roles of these two operations: pixel compression reduces pixel size and expands the FOV while preserving SBP; pixel interpolation further enhances both SBP and FOV through effective pixel size reduction.

**Metasurface phase for FOV expansion**

To achieve holographic display with a near-full-screen field of view, we designed a metasurface composed of 6000×6000 unit cells (pitch: 249.3 nm × 249.3 nm) to reconstruct a uniform light field measuring 95.1 mm × 95.1 mm at a distance of z = 8.39 mm (Fig. 3a). Light propagation was simulated using the angular spectrum method, whereby wavevector components are obtained via a two-dimensional fast Fourier transform (FFT) of the incident wavefront—a computationally efficient approach. Under the small-angle approximation, patterns defined in Cartesian coordinates can be directly mapped to k-space. However, at large diffraction angles, these patterns exhibit pronounced distortion in k-space compared to their Cartesian representations.

To visualize this coordinate-mapping distortion, we illustrate four squares of varying angular extents in Cartesian coordinates (Fig. 3b). Following conventional small-angle practice, equivalent squares are plotted directly in k-space (Fig. 3c). When these k-space profiles are rigorously transformed back to Cartesian coordinates (Fig. 3d), the resulting shapes reveal progressively severe pincushion distortion as the FOV increases: at 40°×40°, distortion is minimal; by 90°×90°, the square's corners stretch toward infinity; and at 160°×160°, the shape deviates drastically from a true square.

To address this distortion, we introduce a correction framework in the spatial frequency domain via a parallel point-to-point search algorithm. The coordinates of a point in the xy-plane at a propagation distance z are given by:

$$x = \frac{\alpha z}{\sqrt{1-\alpha^2-\beta^2}}, y = \frac{\beta z}{\sqrt{1-\alpha^2-\beta^2}}, \quad (1)$$

where $\alpha$ and $\beta$ are the directional cosines along the x- and y-axes, respectively. A key consideration is the discrepancy between the differential area elements $dxdy$ in Cartesian space and $d\alpha d\beta$ in k-space at large diffraction angles. Using the Jacobian determinant in Supplementary Section 6, we derive the transformation

$$dxdy = \left|\frac{\partial x}{\partial \alpha}\frac{\partial y}{\partial \beta} - \frac{\partial x}{\partial \beta}\frac{\partial y}{\partial \alpha}\right| d\alpha d\beta = \frac{z^2}{(1-\alpha^2-\beta^2)^2} d\alpha d\beta = \frac{z^2}{\gamma^4} d\alpha d\beta, \quad (2)$$

where $\gamma = \sqrt{1-\alpha^2-\beta^2}$ represents the directional cosine along the z-axis. Applying Eq. (1) to transform the Cartesian squares from Fig. 3b into k-space yields the rounded profiles shown in Fig. 3e, where all wavevectors within each square remain bounded by $k_o$. While Eq. (1) governs the coordinate transformation, Eq. (2) enables brightness correction to account for energy conservation. In k-space, the 160°×160° FOV square exhibits pronounced compression at its corners. To compensate and preserve energy distribution, the intensity in corresponding regions (Fig. 3f) is scaled by $\gamma^{-4}$. After this adjustment, the log-scaled intensity attains a maximum value of 4,254 at the four corners, indicating that an equivalent of 4,512 pixels in Cartesian coordinates are compressed into a single pixel in the spatial-frequency domain. Therefore, by formulating an appropriate coordinate transformation and implementing a corresponding brightness correction, we effectively mitigate distortions induced by coordinate mapping at wide angles. This approach facilitates efficient k-space optimization using the fast Fourier transform.

As illustrated in Supplementary Fig. 1a, the phase optimization is performed via gradient descent, initialized with a quadratic phase profile (Supplementary Section 1) to promote uniform initial light dispersion and ensure convergence. The error function is defined as a weighted mean square error between the diffraction pattern intensity and the target distribution in spatial frequency space. In accordance with Eq. 2, a weighting factor of $\gamma^4$ is applied at each coordinate to mitigate over-representation of high-spatial-frequency components. As depicted in Fig. 3g, the weighted root-mean-square error (WRMSE) between simulated and target patterns decreases monotonically with iteration count, while diffraction efficiency experiences a slight decline. After 200 iterations, the resulting expansion phase (Fig. 3h) is experimentally realized using titanium dioxide (TiO$_2$) circular nanopillars (fabrication details in Methods; scanning electron microscopy image in Fig. 3k). By applying two-dimensional interpolation coupled with brightness correction, the k-space diffraction pattern is mapped back to Cartesian coordinates, yielding a relatively uniform distortion-free intensity profile (Fig. 3i).

To further validate the accuracy of the retrieved phase (Fig. 3h), we perform cross-verification using Rayleigh–Sommerfeld diffraction[12,53]. The substantial scale difference between the target image (95.1 mm × 95.1 mm) and the metasurface (1.496 mm × 1.496 mm) precludes the use of FFT-based Rayleigh–Sommerfeld propagation[54] for phase optimization, as it requires identical sampling periods at both planes. This constraint would necessitate an impractical grid size (514,560 × 514,560 samples) at the target plane, exceeding typical computational memory limits. While highly accurate, this method is therefore suitable only for numerical verification, not for iterative optimization.

Simulated intensity distributions across a 160°×160° square region at z=8.39 mm, obtained via the angular spectrum method (Fig. 3i) and the split-step Rayleigh–Sommerfeld diffraction (Fig. 3j, see its simulation details in Supplementary Section 3), show strong agreement. Slightly enhanced speckling in Fig. 3j stems from the finite diffraction distance in the Rayleigh–Sommerfeld model, unlike the infinite-distance assumption underlying the design in Fig. 3i. Experimentally, the metasurface is illuminated with quasi-collimated light, yielding the directional pattern in Fig. 3l. Slight barrel and pincushion distortions arise primarily from deviations in collimation and angle of incident beam, which bring pronounced aberrations at such extreme FOVs. The central hotspot corresponds to unmodulated transmitted light in metasurfaces. The measured total efficiency of the metasurface approaches 45.1% (Supplementary Section 4), which may be underestimated due to partial collection loss at large diffraction angles.

**Retrieval of SLM phase for pixel-interpolation meta-holography**

To design the SLM phase, we develop a modified Gerchberg-Saxton algorithm [55] (Fig. 4a) that incorporates data up/down-sampling processes to align with the pixel-interpolation framework of our system. First, the process begins by applying coordinate transformation and brightness correction (Eqs. 1–2) to the target image—here, a "dragon" pattern (Fig. 4b)—mapping it from

Cartesian to spatial-frequency (k-space) coordinates. Second, the initial phase $\varphi_0$ and incident field $A_0$ are then up-sampled from an M×M grid to N×N, where M and N denote the pixel counts along one dimension of the SLM and metasurface, respectively. This ensures the incident field and metasurface share the same spatial sampling (N×N). The up-sampled field, superposed with the metasurface phase, undergoes forward propagation via a Fourier transform (FT), yielding an updated phase $\varphi_T$ for use in the subsequent backward step. Third, Using $\varphi_T$ together with the pre-compensated target image, we perform inverse propagation (denoted as $FT^{-1}$) to obtain the phase $\varphi_{inv}$, which encodes the target image information. The metasurface phase $\varphi_{META}$ is then subtracted from $\varphi_{inv}$ to update the initial SLM phase as $\varphi_{SLM}$, completing one iteration. Finally, upon convergence, the final SLM phase $\varphi_{SLM}$ is obtained by down-sampling the resulting N×N phase distribution back to the original *M*×*M* SLM resolution (see Supplementary Section 5 for details).

Figure 4c presents the optimized SLM phase after 50 iterations, designed to project a "dragon" pattern (Fig. 4d) measuring 95.1 mm × 95.1 mm at a distance of z = 8.39 mm. Here, a pixel-scaling factor of 3 (i.e., *N/M* = 6000/2000) is applied between the metasurface and the compressed SLM. Experimental validation was carried out using a custom-built optical setup (schematic in Fig. 4e; full details in Methods). The reconstructed image captured at the screen (Fig. 4f) clearly reproduces the intended "dragon" pattern without significant distortion, confirming the effectiveness of our distortion pre-compensation strategy for a hologram with a FOV of 159.4° × 159.2°. A central bright spot, attributable to unmodulated incident light, is also observed.

To clarify the functional contribution of the metasurface, we performed a control experiment in which the metasurface was removed while all other components remained unchanged. Both simulations (Fig. 4g) and experimental results (Fig. 4h) show that the output degrades into a diffuse bright spot, even when the SLM is programmed with the phase map from Fig. 4c. This outcome underscores that the metasurface is indispensable not only for expanding the FOV but also for enabling meaningful image formation.

We further investigated the impact of misalignment between the compressed SLM phase and the metasurface by introducing controlled lateral shifts ($\Delta x$) ranging from 0 to 400 μm. As shown in Fig. 4i, the reconstructed dragon pattern remains discernible but becomes progressively cropped with increasing offset. Notably, the holographic image retains its structural integrity even under substantial displacement and does not vanish, demonstrating a favourable tolerance to misalignment that supports practical implementation. This advantage originates from the regularly distributed phase of metasurfaces so that the lateral misalignment leads to only a small shift of the reconstructed images and will not destroy the image quality severely. In comparison, other methods based on random distributed phase are very sensitive to the misalignment between the SLM phase and the fabricated masks[23,51], thereby increasing the experimental difficulty in achieving the expected images.

Notably, our approach imposes no constraints on the selection of target images—a distinct advantage over conventional dynamic meta-holography systems, which typically support only a limited set of pre-designed patterns. For any given target image, the corresponding SLM phase can be optimized using the algorithm outlined in Fig. 4a. Furthermore, dynamic switching between different images is achieved by simply updating the SLM phase in real time.

**Large-FOV meta-holographic images**

To illustrate the advantage of our approach over the pixel-compression-only approach, we designed a holographic image with a field of view (FOV) of 160°×160°, approaching the theoretical expansion limit of the fabricated metasurfaces. The experimentally captured result (Fig. 5a) confirms a FOV of ~159°×159°, though some nonuniformity is observed due to insufficient sampling in the spatial-frequency domain, suggesting that denser sampling would further improve reconstruction fidelity.

In a control experiment, we removed the metasurface from the optical setup (Fig. 4e), leaving only the pixel-compression architecture. Based on the compressed SLM pixel size (~0.748 μm), the theoretical maximum FOV is 44°. However, constrained by the numerical aperture of lens $L_4$ (focal length: 30 mm, diameter: 1 inch), the SLM-only hologram was redesigned for half of this maximum FOV. The resulting pattern (Fig. 5b) exhibits a measured FOV of ~22.3°×22.3°, see more experimental details in Supplementary Section 7. By comparison, our metasurface-assisted system achieves a FOV enhancement of approximately 7×7 relative to the pixel-compression-only case. This result represents the largest FOV reported to date among dynamic holographic systems[15,16,18-22,24] (see Fig. 1d for comparative analysis), underscoring the critical role of our method in advancing full-screen holographic displays.

**Large-FOV meta-holographic movies**

To demonstrate the dynamic capabilities of our system, we precomputed SLM phase patterns for each frame of a video sequence and addressed them sequentially via their corresponding frame indices. The field of view (FOV) for all frames was set to 157.5°×157.5° to minimize distortion after pre-compensation. Using the optical setup in Fig. 4e, dynamic holographic video was successfully projected onto the screen. Selected frames from the reconstructed sequence are shown in Fig. 5c, depicting a large "dolphin" swimming across the wide angular range. The full video, recorded continuously, is available as Supplementary Movie 1—to our knowledge, the first holographic video demonstrated with such a high FOV.

The frame rate was characterized by toggling between two distinct holographic frames—one with high intensity and the other with low intensity—at the same screen position (insets, Fig. 5d). A photodetector placed at this location recorded a near-periodic electrical signal (Fig. 5d), where high- and low-amplitude levels correspond to the first and 26th frames, respectively. Over a one-second interval, 30 cycles of high and low signals were observed. Fast Fourier transform (FFT) analysis of the temporal signal confirms a fundamental frequency of 30 Hz (Fig. 5e). Since each cycle comprises two distinct frames, the system operates at a video frame rate of 60 Hz, exceeding the typical temporal resolution threshold of the human eye (~24 Hz).

Compared to previously reported dynamic meta-holograms[46,48], our pixel-interpolation approach uniquely combines a near-full-screen FOV, high refresh rate, and support for unlimited image content. We summarize key performance metrics—including frame rate, effective pixel size, and theoretical image capacity—for several dynamic meta-holographic systems in Fig. 5f [24,46,48-51,56,57]. The comparison highlights that existing systems typically compromise on at least one of these metrics, whereas our method satisfies all essential requirements for practical holographic display, occupying the optimal region in the parameter space (orange area, Fig. 5f). Note that, although our previous approach also meets these metrics[24], the FOV is still limited to 38.9°×50.3°, which is far below 160°×160° in our current work. Therefore, these capabilities make the platform particularly suitable for applications in virtual and augmented reality, where high spatial and temporal performance is critical.

**Discussion**

The FOV in our dynamic meta-holography system could be further extended by engineering metasurface phases with stronger wavefront-expanding properties. Given the subwavelength architecture of metasurfaces, the theoretical FOV limit approaches 180°×180°. For practical deployment, future efforts should focus on increasing the frame rate and enabling multi-color operation. The present video refresh performance can be significantly improved by adopting high-performance graphics processing units and high-speed data interfaces, especially since the Holoeye GAEA SLM natively supports refresh rates up to 60 Hz. Realizing full-color display will require metasurface redesign to mitigate chromatic aberration, along with utilization of the SLM's multi-color mode for time-sequential channel rendering.

In summary, we have realized a pixel-interpolation-assisted dynamic meta-holography platform that reconstructs arbitrary images with an ultra-wide field of view. Our approach synergistically combines the subwavelength pixelation of metasurfaces with the dynamic programmability of SLMs. To address distortion arising from spatial-frequency coordinate mapping, we introduced a tailored coordinate transformation and brightness compensation framework in $k$ space. We further developed a modified Gerchberg–Saxton algorithm incorporating up-sampling and down-sampling operations to implement pixel interpolation. This methodology enables the design of SLM phase patterns that—when optically combined with a metasurface—generate holographic images with exceptionally wide FOVs. Experimentally, we demonstrated a holographic display with a FOV of 157.5°×157.5° at a refresh rate of 60 Hz, thereby advancing the prospects of high-dynamic-range, large-FOV holography for virtual reality and assisted driving systems.

**Methods**

**Numerical simulations**. In this work, the angular spectrum method is implemented by using a Fourier transform of the incident field. The outputted field is calculated in the $k$-space that can be expressed approximately in terms of the ratio of the lateral position at the target plane to the propagation distance. Under the condition obeying the sampling theorem, the sampling interval at the target plane can be customized arbitrarily in principle to avoid the aliasing effect.

The verification of the phase designed by the Frourier trasnform is implemented via the Rayleigh-Sommerfeld diffraction[12,53]. After the designed SLM phase is interpolated by the metasurfaces, the resulting phase has the subwavelength pixels. Light from each pixel is considered as an ideal point source that can be described by using an analytical diffraction kernel in Rayleigh-Sommerfeld diffraction. Thus, the diffraction field at the target plane can be taken as a superposition of light from all the pixels that are modulated by the corresponding phase. Because the diffraction kernel and the resulting phase are given, we can calculate the rigorous electric fields at arbitrary positions of the target plane without the sampling issues in the FFT-based approaches. The sampling interval at the target plane can be customized arbitrarily due to the analytical diffraction kernel. However, this weighted-summation approach is quite time-cost to calculate the Rayleigh-Sommerfeld diffraction. Due to the symmetry of the square, only the diffraction pattern in the first quadrant is calculated. In this work, the resulting phase has a sampling amount of 6000×6000 (the sampling interval is 249.3 nm) and its diffraction pattern located at a propagation distance of $z$= 8.39 mm is sampled with 252000× 252000 in the first quadrant. It takes 2.7 hours to obtain the first quadrant of the diffraction pattern Fig. 3j in a computer (Intel Core CPU i7-12700 @ 2.1G Hz, RAM 32GB). The simulated result is provided in Fig. 3j, which shows good agreement with the experimental pattern in Fig. 3l. In despite of its accuracy, the summation-based approach is inefficient in optimizing the holographic phase.

**Design and fabrications of metasurfaces**. To realize the metasurfaces, we use circular-shape $TiO_2$ nanopillars with varying diameters on a glass (BF33) substrate. To obtain sufficient phase modulation at the operating wavelength of 561 nm, the heights of the nanopillars are fixed at 613 nm. The sketch of a single unit cell with a pitch of 249.3 nm ×249.3 nm is shown in Supplementary Fig. 2a. To simulate optical properties of each nanopillar, the finite-difference time-domain method is used here with a periodic boundary condition along $x$ and $y$ directions and perfect matching layers along the $z$ direction. The simulated transmission and the phase delay of these nanopillars are presented in Supplementary Fig. 2b, which shows a phase modulation of $2\pi$ via changing the diameter of the nanopillars while the transmission maintains over 90% for most diameters.

The metasurface was fabricated via a sequence of nanofabrication steps beginning with electron-beam lithography. A $TiO_2$ film of 613 nm thickness deposited on a substrate was spin-coated with a positive-tone electron-beam resist (AR-P 6200) and soft-baked. Exposure was carried out using a 100 kV electron-beam lithography system (JEOL JBX 6300FS). After development, the sample underwent a post-exposure bake to eliminate residual moisture. A 10 nm chromium hard mask was subsequently deposited by electron-beam evaporation. Lift-off was performed to pattern the mask, followed by etching of the $TiO_2$ layer. Finally, the remaining chromium was removed using a selective wet etchant.

**Experimental setup**. Figure 4e sketches the optical setup to characterize the pixel-interpolation-based dynamic meta-holograms. A laser with a wavelength of 561 nm is expanded by using a telescope system composed of two spherical lenses $L_1$ and $L_2$ so that the incident beam size can match that of the active region (2000×2000 pixels) of a reflective SLM. After carrying the SLM phase, the light is scaled down by using another telescope system (Lenses $L_3$ and $L_4$) to realize pixel-pitch compression from 3.74 μm×3.74 μm (SLM's original pixel pitch) to 0.748 μm×0.748 μm (compressed pixel pitch). Thus, the compressed SLM phase is located at the rear plane of the lens $L_4$. To interpolate the metasurface phase into the compressed SLM phase, we just put the fabricated metasurfaces at the rear plane of the lens $L_4$ by using high-precision three-dimensional stages.

## Acknowledgements

This work is supported by the National Key Research and Development Program of China (2022YFB3607300), the National Natural Science Foundation of China (Grant Nos. 62322512, 62225506, 62422506, 62505308, 12474383 and 12134013), the Fundamental Research Funds for the Central Universities (WK2030000108, WK2030000090), CAS Project for Young Scientists in Basic Research (Grant No.YSBR-049). K. H. thanks the support from the University of Science and Technology of China's Centre for Micro and Nanoscale Research and Fabrication. D. Z. thanks the China Postdoctoral Science Foundation (2023M743364) and Anhui Natural Science Foundation (2508085QA010). Z.-L. D. thanks the Guangdong Provincial Quantum Science Strategic Initiative (GDZX2406004), the Guangdong Basic and Applied Basic Research Foundation (2022B1515020004), and the Guangzhou Science and Technology Program (2024A03J0465, 2025A04J5776). The numerical calculations were partially performed on the supercomputing system at Hefei Advanced Computing Center and the Supercomputing Center of the University of Science and Technology of China.

## Author contributions

K. H. conceived the idea. R. X., F. L., Q. D. and Y. L. conducted the hologram design. F. L., R. X., Q. D., X. L. and Z. D. designed the metasurfaces. F. L. and D. Z. fabricated the samples. R. X., F. L., Y. L. and D. Z. built the experimental setup and characterized the samples. R. X., F. L., K. H. and F. S. visualized and analyzed the data. K. H., R. X., F. L., D. Z. and Z. D. wrote the manuscript. K. H., Z. D. and D. Z. supervised the project. All the authors discussed the results.

## Competing interests

The authors R. X., Y. L., X. L., D. Z. and F. S. claim no competing interests. The authors Z. D., F. L., Q. D. and K. H. declare the following competing interests. Z. D., F. L., Q. D. and K. H.. have filed two patent application related to this work through Jinan University and the University of

Science and Technology of China. The first patent (Z. D., F. L., Q. D., Z. W., M. H. and K. H., "A metasurface-based dynamic color holographic display method, system, device and medium", patent No. ZL202410162813.4 (2024)) has been granted. This patent applied by Jinan University and University of Science and Technology of China refers to the design method and physical architecture of metasurface-based SLM for large-FOV dynamic display. The second patent (Z. D., F. L., Q. D., Z. W., M. H. and K. H., "A phase retrieval method for spatial-field-based metasurface-based large-FOV holography", patent No. ZL202410162953.1 (2024)) has been granted. This patent applied by Jinan University and University of Science and Technology of China refers to the phase retrieval in metasurface-based SLM for large-FOV dynamic display.

**Figures and Figure Captions**

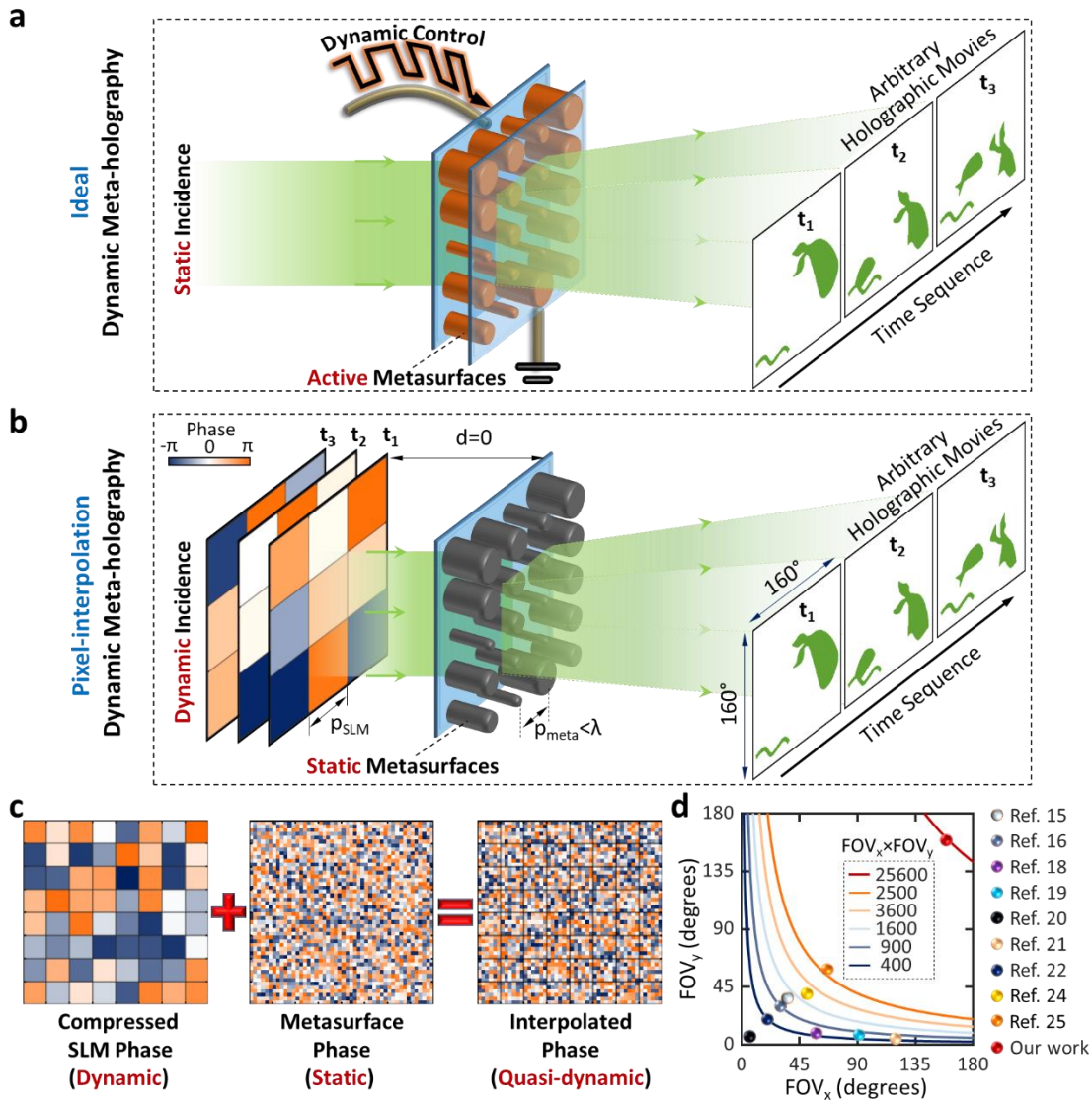

**Figure 1. Principle of the pixel-interpolated dynamic meta-holography**. (a) Ideal dynamic meta-holography by controlling optical properties of each nanostructure actively. It works in a fashion of static incidence and active metasurfaces for dynamic reconstruction of holographic images in a time sequence. However, the challenging active metasurfaces with high reconfigurabilities remain unavailable now. (b) Our proposed pixel-interpolated dynamic meta-holography find a new way to employ static metasurfaces and dynamic incidence. The active incidence releases the challenge of realizing active metasurfaces. The static metasurfaces work as a functionalized device that offers large FOV. It can realize the same functionalities of ideal dynamic meta-holography in reconstructing arbitrary images. (c) Sketch for the pixel-interpolation

operation in our proposal. The dynamic SLM phase with large pixel pitches is interpolated optically by different metasurface phases with subwavelength pixel pitches. The interpolated phases have both quasi-dynamic and subwavelength pixel pitches for reconfigurable large-FOV holograms. **(d)** A comparison of full-screen FOVs among the reported dynamic holograms for reconstructing arbitrary images. For the works reporting only one-direction FOV[18,19,21], the FOVs along the other direction are extracted from their operating wavelengths and the pixel pitches. It shows the largest full-screen FOV for our proposed dynamic meta-holography.

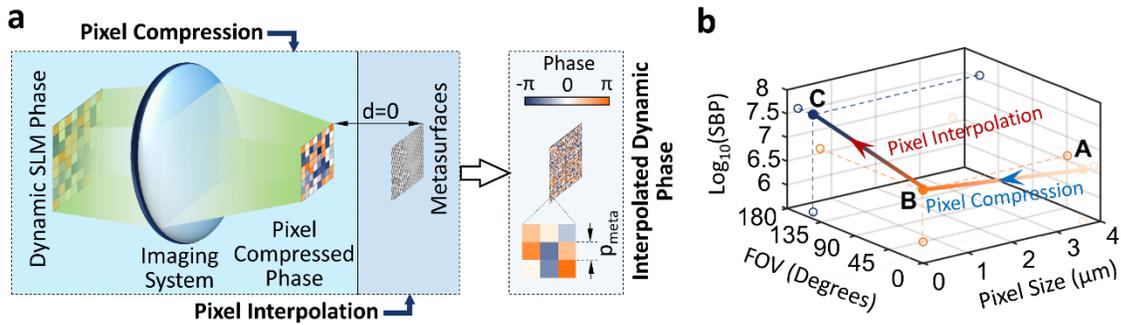

**Figure 2. Strategy for practical implementation of the pixel-interpolated dynamic meta-holography**. **(a)** Sketch for optical realization of pixel compression and pixel interpolation. The dynamic SLM phase is compressed by an imaging system and projected on the imaging plane. The metasurfaces are located at the imaging plane to realize pixel interpolation optically because the distance between the compressed SLM phase and the metasurface phase can be tuned mechanically to be $d=0$ in the experiment. **(b)** Spatial-bandwidth product (SBP), FOV and efficient pixel size during the operations of pixel compression and pixel interpolation in the practical pixel-interpolated dynamic meta-holography.

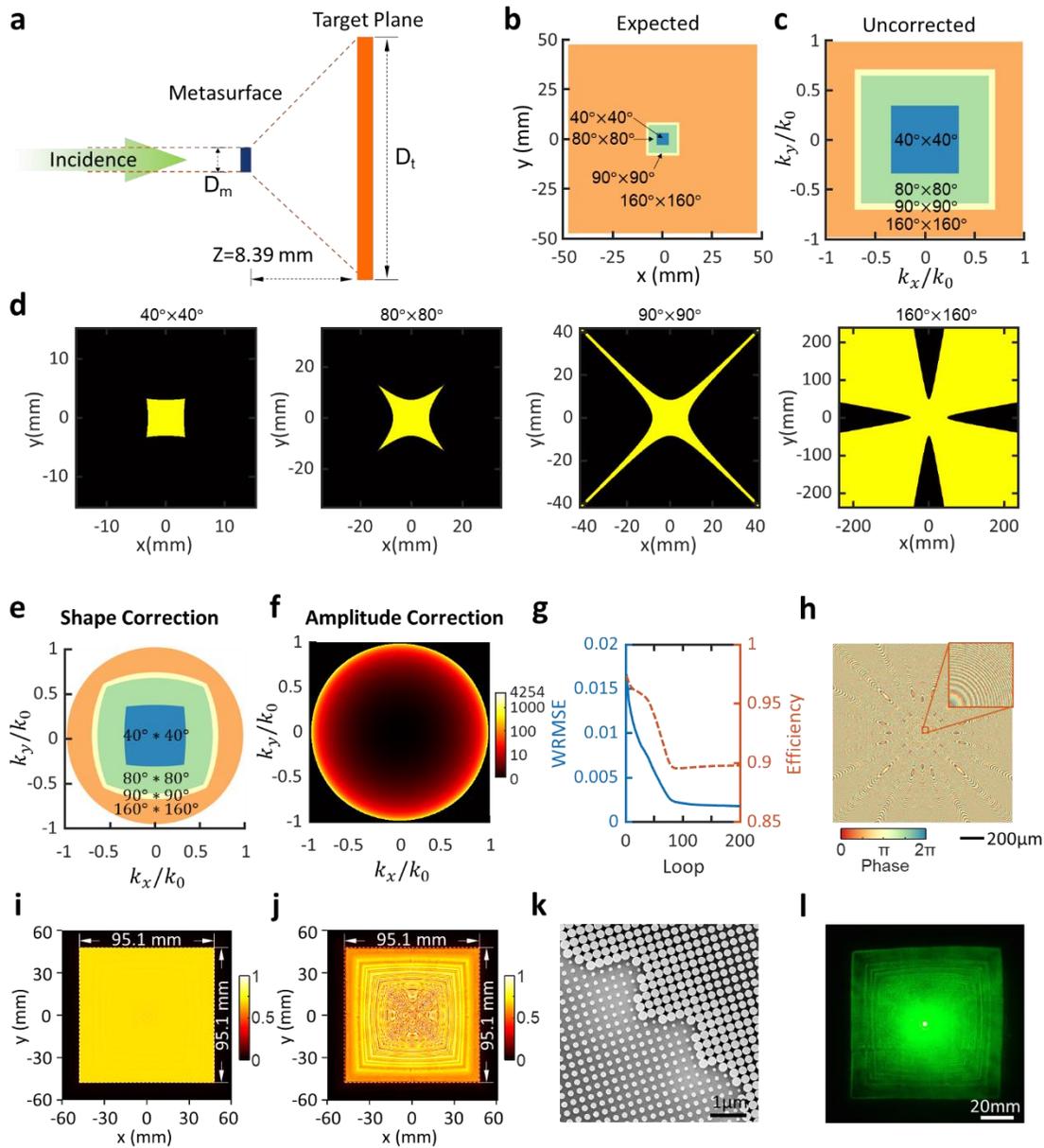

**Figure 3. Designing procedure of the FOV expansion metasurface for dynamic hologram**. (**a**) Mechanism for designing and characterizing the FOV expansion metasurface. It is implemented by illuminating the metasurfaces (1.496 mm×1.496 mm) with an incident plane wave to generate a uniformly distributed square-shape field (95.1 mm×95.1 mm) with expanded FOV at the propagation distance of $z$=8.39 mm. The propagation of light is approximated by using the Fraunhofer diffraction. (**b**) Four squares in Cartesian coordinates with different full FOVs. Four distortion corrected squares in the spatial frequency space with different FOVs. (**e**) Four squares in the $k$-space (**c**) mapped to the Cartesian coordinate system. (**f**) The brightness correction factor of the 160° FOV square, which equals to the fourth power of the spatial cosine $\gamma$. **(g)** Weighted root-mean-square error (WRMSE) and diffraction efficiency of the simulated patterns vs the iteration steps (from 0 to 200) of metasurface phase optimization. **(i)** Simulated intensity profile from the designed metasurface phase **(h)** when illuminated by a plane wave. **(j)** Simulated intensity profile from the designed metasurface phase **(h)** by Rayleigh-Sommerfeld diffraction method. **(k)** SEM image (with a titling view) of our fabricated metasurfaces. **(l)** Experimentally captured image at the designed distance of z= 8.39 mm from the metasurface. Because the size of the reconstructed pattern exceeds the effective area of most CCD or CMOS detectors, we used a camera to directly capture the metasurface sample. The hot spot at the center of the sample denotes the unmodulated incident light on the metasurface.

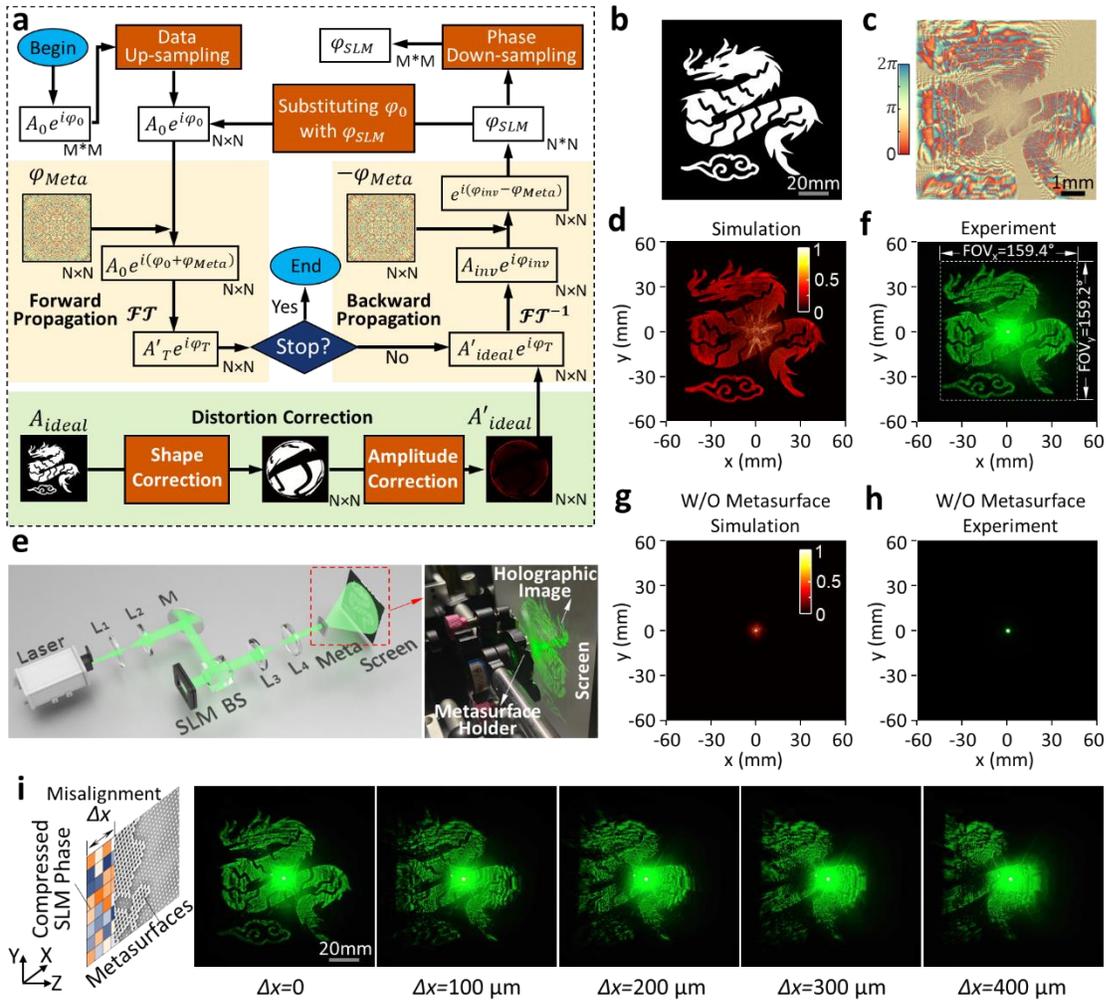

**Figure 4. Pixel-interpolation meta-holograms. (a)** Flowchart of designing the SLM phase for pixel-interpolation meta-holograms. The up-sampling operation is implemented by dividing one SLM pixel into $(N/M)^2$ metasurface pixels with the same phase, where $N$ is the integer times of $M$ in our proposal. The down-sampling operation is realized by taking one out of every $N/M$ metasurface pixels, thereby making the $N \times N$ metasurface phase profile into a $M \times M$ phase profile for SLM. In this work ($N/M=3$), each phase at the up-left corner of every 3×3 meta-pixel square is employed to match the $M \times M$ SLM phase. **(b)** Target holographic image with a "dragon" pattern. **(c-d)** Designed SLM phase **(c)** for generating the simulated image **(d)**. In the hologram design, the distortion pre-compensation is used here to avoid the distorted pattern. **(e)** Experimental setup for characterizing the holographic image and the image of the metasurface diffraction. The inset shows the metasurface device and the screen in holographic display in our laboratory environment. **(f)** Experimentally captured image without obvious distortion. **(g-h)** Simulated **(g)** and experimental **(h)** patterns by removing only the metasurfaces in **(e)**. It is taken as a control experiment to validate the essential role of the metasurfaces in the extremely wide FOV hologram reconstruction.

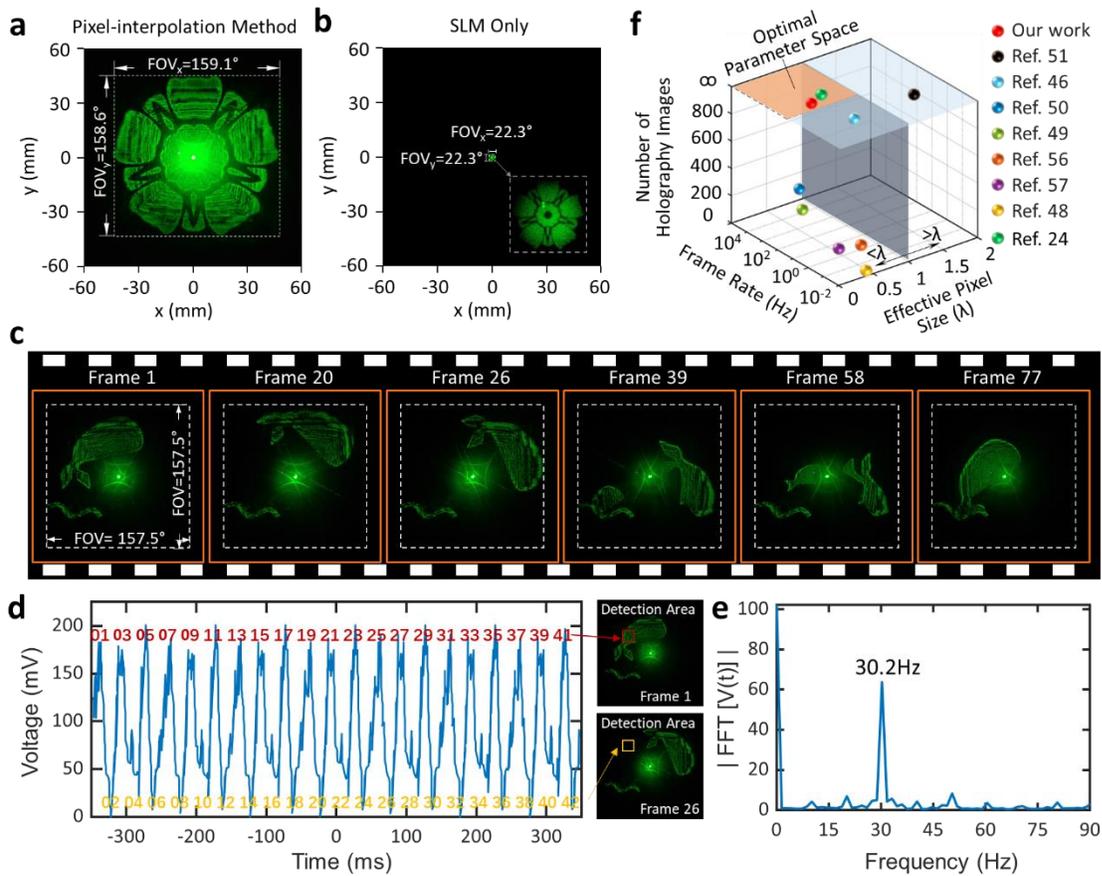

**Figure 5. Large-FOV and dynamic features of our pixel-interpolated meta-holography. (a-b)**

Comparison of experimentally reconstructed images by using our pixel-interpolation method **(a)** and the SLM only **(b)**. It indicates that our method can yield a FOV of 159.1°×158.6° , largely surpass the traditional SLM-only method which create the same pattern with a FOV of 22.3°×22.3°. Figures **(a)** and **(b)** were captured by the camera at the same propagation distance (removing the metasurface slightly affects the actual working distance due to the refractive index of the substrate). A more general comparison of FOVs among different dynamic holograms is provided in Fig. 1d for a better observation. **(c)** Selected frames in our demonstrated holographic video with a near full-screen FOV of 157.5 °×157.5 °. **(d)** Time-dependent voltage signals of optical detector recording optical fields of two switchable frames. The high and low voltages are relative to the $1^{st}$ and $26^{th}$ frames, respectively. **(e)** Fourier transform of the temporal signals in (d) for determining its period, which turns out to be 30. It doubly confirms that 60 frames per second can be achieved in our system. **(f)** Detailed comparison about optical performances (*i.e.*, the largest number of holographic images, effective pixel size and frame rate) of various dynamic meta-holograms. For the holographic display, the optimal parameter space (the orange part) has unlimited image number, subwavelength-scale effective pixel size and high (>24 Hz) frame rate. As the works [46,51,56] that do not provide refreshing rate, we assume that their refreshing rates are 1 for a better comparison.

Supplementary Information for

# A 160°×160° Dynamic Holographic Meta-projector

*Feng-Jun Li[1,#], Ruixing Xia[2,#], Qianmei Deng[1,#], Yuze Lu[2,#], Xiangping Li[1], Fangwen Sun[2], Dong Zhao[2,\*], Zi-Lan Deng[1,\*], Kun Huang[2,3,\*]*

[1]Guangdong Provincial Key Laboratory of Optical Fiber Sensing and Communications, Institute of Photonics Technology, College of Physics & Optoelectronic Engineering, Jinan University, Guangzhou, 510632, China

[2]Department of Optics and Optical Engineering, University of Science and Technology of China, Hefei 230026, China

[3]National Key Laboratory of Opto-Electronic Information Acquisition and Protection Technology, Anhui University, Hefei, 230601, China

[#] F. L., R. X., Q. D. and Y. L. contributed equally to this work.
[*] Corresponding authors: D. Z. (zd10375@ustc.edu.cn), Z.-L. D. (zilandeng@jnu.edu.cn), or K. H. (huangk17@ustc.edu.cn)

## Table of Contents

## ■ Supplementary Section 1. Design of metasurface phase for field-of-view (FOV) expansion

Metasurfaces are born with the capabilities of extending the FOV due to its subwavelength pixel feature. In principle, the lens phase could provide the FOV expansion capability. However, off-axis aberration of the lens will dominate the imaging process for a large-FOV so that the reconstructed image might have poor quality. Therefore, in this work, we design a FOV expansion metasurface by using a gradient descent algorithm [1].

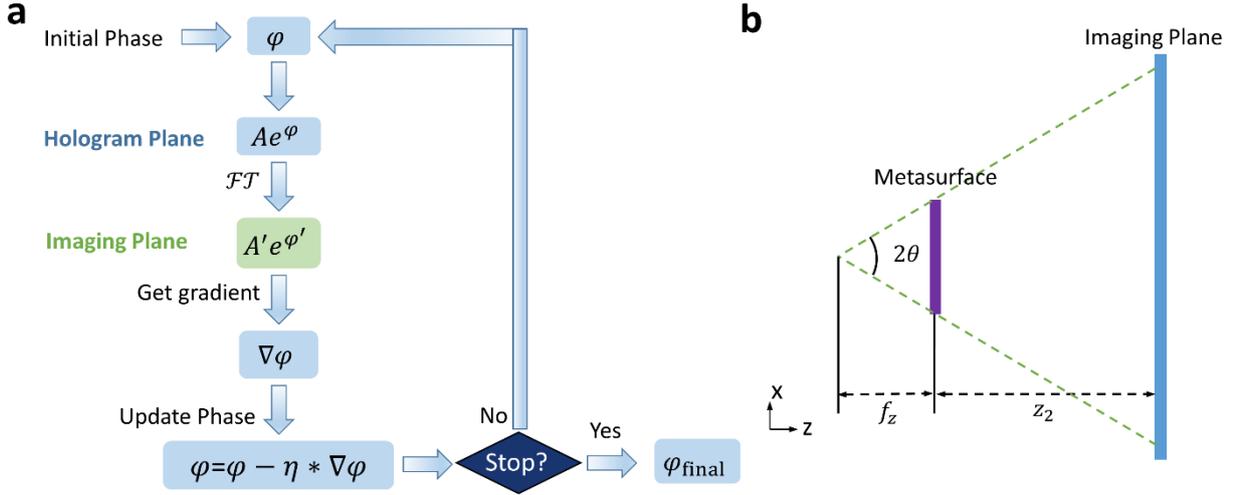

**Supplementary Figure 1. Optimization of metasurface phase for FOV expansion.** (**a**) Flowchart of the modified GS algorithm to optimize metasurface phase. (**b**) Sketch of optical path for designing the metasurface phase.

The flow chat of the optimization algorithm is shown in Supplementary Fig. 1a. An initial quadratic phase $\varphi_h$ is first multiplied with amplitude $|A|$ on the hologram plane, and the resulting complex amplitude is set as the input of the iterative process. The initial phase $\varphi_h$ takes the form of $\varphi_h(x,y) = e^{ik(f_z - \sqrt{x^2+y^2+f_z^2})}$, where $k$ is the wavevector in vacuum, $f_z$ is the focal length. As sketched in Supplementary Fig. 1b, the focal length is determined by the geometric relationship. The size of the metasurface is set as 1.496×1.496 mm. We obtain $f_z = \frac{-L}{2*tan(80°)} \approx -\frac{1.496}{2*5.67} mm \approx -0.132 mm$. The minus sign of the focal length means that it's a concave lens.

In forward propagation, we use single fast-Fourier-transform (FFT) to calculate spatial frequency components in the *k*-space. The loss function is defined as the weighted mean-square error (WMSE) between the intensity in the *k*-space and the target intensity in the *k*-space, that is

$$L = \sum_{p=0}^{M-1} \sum_{q=0}^{N-1} \alpha_{pq} (I_{pq} - I_{t,pq})^2 \qquad (S1)$$

where $\alpha_{pq} = \gamma^{-4}$ is the weight factor in which $\gamma = \sqrt{1 - \alpha^2 - \beta^2}$ is the directional cosine along the z-axis, $I_{t,pq}$ is the target intensity in the *k*-space. Following Eq. 2, the weight factor is set equal to the fourth power of the gamma factor ($\gamma^4$) in the *k*-space to prevent over-representation by points with high spatial frequencies. According to the chain rule, we get

$$\frac{\partial L}{\partial \varphi_{pq}} = \sum_{k,l} \frac{\partial L}{\partial I_{kl}} \cdot \frac{\partial I_{kl}}{\partial \varphi_{pq}} \qquad (S2)$$

The first term is straightforward, as $L$ is a simple quadratic function of $I_{kl}$,

$$\frac{\partial L}{\partial I_{kl}} = 2\alpha_{kl}(I_{kl} - I_{t,pq}). \qquad (S3)$$

The intensity in the *k*-space is calculated as the squared magnitude of the complex electric field $E_{kl}$, which is the 2D-discrete Fourier transform (DFT) of the incident field $U_{mn}$,

$$E_{kl} = \sum_{mn} U_{mn} \phi_{klmn}, \qquad (S4)$$

where $\phi_{klmn} = \exp\left(-i2\pi \left(\frac{mk}{M} + \frac{nl}{N}\right)\right)$ is the DFT kernel. By using the product rule, we obtain

$$\frac{\partial I_{kl}}{\partial \varphi_{pq}} = \frac{\partial E_{kl} E_{kl}^*}{\partial \varphi_{pq}} = E_{kl} \frac{\partial E_{kl}^*}{\partial \varphi_{pq}} + E_{kl}^* \frac{\partial E_{kl}}{\partial \varphi_{pq}}. \qquad (S5)$$

The term $\frac{\partial E_{kl}}{\partial \varphi_{pq}}$ in Eq. S5 follows

$$\frac{\partial E_{kl}}{\partial \varphi_{pq}} = \frac{\partial \sum_{mn} U_{mn} \phi_{klmn}}{\partial \varphi_{pq}} = U_{pq} \frac{\partial \phi_{klpq}}{\partial \varphi_{pq}} + \phi_{klpq} \frac{\partial U_{pq}}{\partial \varphi_{pq}} = U_{pq} * 0 + \phi_{klpq} \frac{\partial U_{pq}}{\partial \varphi_{pq}} = i\phi_{klpq} E_{kl}. \qquad (S6)$$

The conjugate term of Eq. S6 is

$$\frac{\partial E_{kl}^*}{\partial \varphi_{pq}} = \left(\frac{\partial E_{kl}}{\partial \varphi_{pq}}\right)^* = -iU_{pq}^*\phi_{klpq}^*. \tag{S7}$$

Substituting Eqs. S6-7 back to Eq. S5, we obtain

$$\frac{\partial I_{kl}}{\partial \varphi_{pq}} = iU_{pq}\phi_{klpq}E_{kl}^* - iU_{pq}^*\phi_{klpq}^*E_{kl} = -2Im\left(U_{pq}\phi_{klpq}E_{kl}^*\right). \tag{S8}$$

Substituting Eqs. S3 and S8 back into Eq. S1, we get

$$\frac{\partial L}{\partial \varphi_{pq}} = -4\sum_{k,l}\alpha_{kl}(I_{kl} - I_{t,pq})Im\left(U_{pq}\phi_{klpq}E_{kl}^*\right). \tag{S9}$$

To enable efficient computation, the double sum is recognized as an inverse DFT. By defining an assistant function,

$$H_{kl} = \alpha_{kl}(I_{kl} - I_{t,pq})E_{kl}^*. \tag{S10}$$

The phase gradient can be simplified as

$$\frac{\partial L}{\partial \varphi_{pq}} = -4Im\left[U_{pq} \cdot (\mathcal{F}\{H_{kl}\})(p,q)\right], \tag{S11}$$

where $\mathcal{F}$ is a DFT. This algorithm is highly computationally efficient, requiring only two 2D FFTs per iteration to get the phase gradient of all points in the incident plane: one for calculating the light intensity ($I_{kl}$) in $k$-space and the other for computing the phase gradient. In this optimization, the phase gradient is first normalized by its maximum absolution value and then multiplied by a learning rate ($\eta$) that gradually decreases throughout the optimization.

$$\varphi = \varphi - \eta * \frac{\partial L}{\partial \varphi}\bigg/\left|\left(\frac{\partial L}{\partial \varphi}\right)\right|_{max}, \tag{S12}$$

$$\eta = 0.4e^{-\frac{m}{100}}, \tag{S13}$$

where $1 \leq m \leq 200$ is the iteration count.

◾ **Supplementary Section 2. Design of nanostructures in metasurfaces**

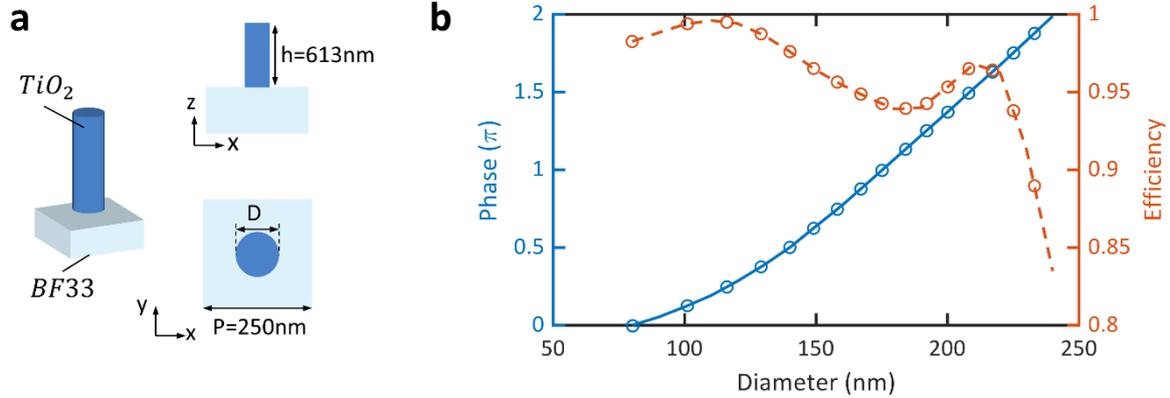

**Supplementary Figure 2. Design detail of the FOV expansion metasurface. (a)** Sketch of one unit cell in the dielectric metasurface made of circular TiO₂ nanopillars on a glass substrate. **(b-c)** Simulated transmission **(b)** and corresponding phase **(c)** modulation with the normal incidence of a plane wave.

In this work, we utilize dielectric metasurface made of circular TiO$_2$ nanopillars on a glass substrate, which has low absorption at the operating wavelength. The sketch of one unit cell of the metasurface is shown in Supplementary Fig. 2a, where the different views of the unit cell are shown for a better observation of our employed metasurfaces. The unit-cell of the metasurface is composed of circular titanium dioxide (refractive index $n$=2.3 at the wavelength of 561 nm) nanopillars on BF33 (refractive index $n$=1.46) substrate, with height $H = 613$ nm, period $P = 250$ nm, and diameter $D$ varying between 80 nm and 233 nm.

The numerical simulation about optical properties of these nano-pillars is implemented by the finite-difference time-domain (FDTD) method[2]. The simulated transmission and phase profiles of nanopillars with different dimensions are presented in Supplementary Fig. 2b. One can find that the optical transmission is over 90% for most nano-pillars. Overall, a full phase modulation of $2\pi$ is achieved for these nano-pillars with the diameters ranging from 80 nm to 233 nm, thus it guarantees the require phase for FOV expansion.

◾ **Supplementary Section 3. Simulating diffraction by the Rayleigh-Sommerfeld theory**

Since our work is focused on large-FOV holographic generation, the employed angular spectrum method (ASM) at infinite distance cannot give exact simulation of the large-FOV diffraction pattern from a metasurface phase because it leads to the distorted patterns in practical experiment, as discussed in main texts. Here, we give the rigorous simulation of the large-FOV diffraction pattern by using Rayleigh-Sommerfeld (R-S) theory[3], which provides more accurate solutions of the optical propagation in freespace and can be derived directly from Maxwell equations.

As the optical parameters (small pixel pitch of 250 nm×250 nm in metasurfaces, long propagation distance of 8.39 mm, the *xy*-axis simulation region of 125.7 mm×125.7mm) are cross-scale, it is difficult to simulate the diffraction pattern by using FFT-based direct integral according to R-S theory. A step-by-step FFT is used to calculate the diffraction image at the target plane. Due to the symmetry of the square, only the diffraction pattern in the first quadrant is calculated. The incident plane contains 6000×6000 pixels while the target plane contains 252000×252000 pixels in the first quadrant. FFT-accelerated 2D matrix convolution requires padding to a size of 351,999 × 351,999, which corresponds to the sum of both side pixel numbers minus one. To conserve computer memory, we partition the target plane matrix into 21 × 21 tiles. Each tile measures 12,000 × 12,000 pixels. By keeping the size of a single 2D-FFT matrix within the 17,999 × 17,999 pixels, 32 GB of memory is sufficient to meet the computational requirements. To reduce storage requirements, the simulated intensity matrix by R-S theory is down-sampled by 100:1 (averaging over 100×100 pixel blocks). It takes 2.7 hours to obtain the first quadrant of the diffraction pattern of Fig. 3j in a computer (Intel Core CPU i7-12700 @ 2.1G Hz, RAM 32GB). The simulated result in Fig. 3j shows good agreement with the experimental pattern in Fig. 3l.

■ **Supplementary Section 4. Characterizing efficiency of the metasurface**

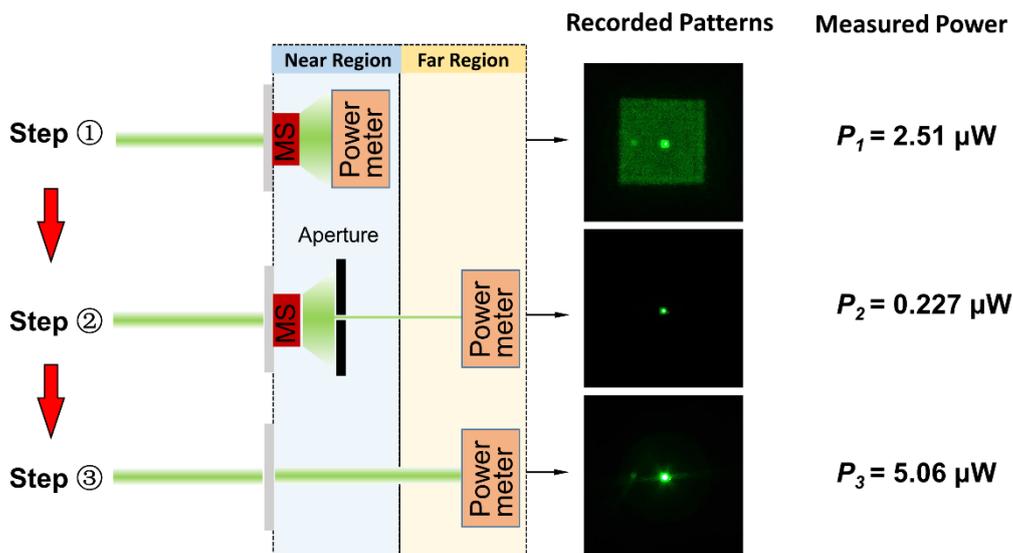

**Supplementary Figure 3. Steps of characterizing the experimental efficiency of the metasurface.**

As shown in Supplementary Fig. 3, three steps are used to calculate the efficiency of the fabricated metasurface. First, the metasurface is illuminated by the collimated light with a size smaller than the metasurface with the help of the SLM. The metasurface's diffraction angular range reaches 160° × 160°, which exceeds the detection range of our power meter probe. When light illuminates the power meter probe at large oblique angles, it results in an underestimation of the power measurement. A 0.15 × 0.15 mm square beam (1/10 side length of the metasurface) illumination is generated by programming a specific phase pattern on the SLM. The central 200 × 200 pixels of the SLM is set to zero, while a buffer region with an outwardly divergent radial phase gradient (8 pixels per 2π) is applied to the surrounding area. The square transmitted light records a total power of $P_1$ = 2.51 μW including a central hot spot.

Second, we adopt a small aperture to measure the power of the central hot spot. The central spot is attributed to unmodulated light transmitting the metasurface which is at a power of $P_2$ = 0.227 μW. To avoid overexposure, the second image at the central-right panel of Supplementary Fig. 3 was captured with a much shorter exposure time compared to the first image (0.02s vs 0.5s), which results in the spot appearing dimmer. Thus, we can obtain the power of the signal light by using $P_1$ – $P_2$=2.283 μW.

Finally, to measure the total power incident on the metasurface, we move the metasurface outside the incident-light region, so that we can record the total power (measured with $P_3$ = 5.06 mW) incident on the metasurface substrate by using the power meter in the far region. The recorded pattern (at the bottom-right panel of Supplementary Fig. 3) shows a square shape of the incident light. As a result, the total efficiency of the metasurface can be calculated as $\frac{P_1 - P_2}{P_3} \approx 45.1\%$. Although it is lower than the simulated efficiency of ~90% due to the imperfect fabrication and unavoidable measurement noise, the experimental efficiency of 45% is sufficiently high at the operating green wavelength of 561 nm. Considering illumination with a 0.15×0.15 mm rectangular beam, the FOV remains as high as 59°. Also, the metasurface is not designed for a beam of this size, a few lights are diffracted into the 59–160° FOV range, resulting in diffraction loss. When a beam with such large diffraction angle (59° full FOV) illuminates the power meter probe, the measured power may be underestimated, potentially leading to an underestimation of the efficiency mentioned above.

◼ **Supplementary Section 5. Pixel match between compressed phase and metasurface phase**

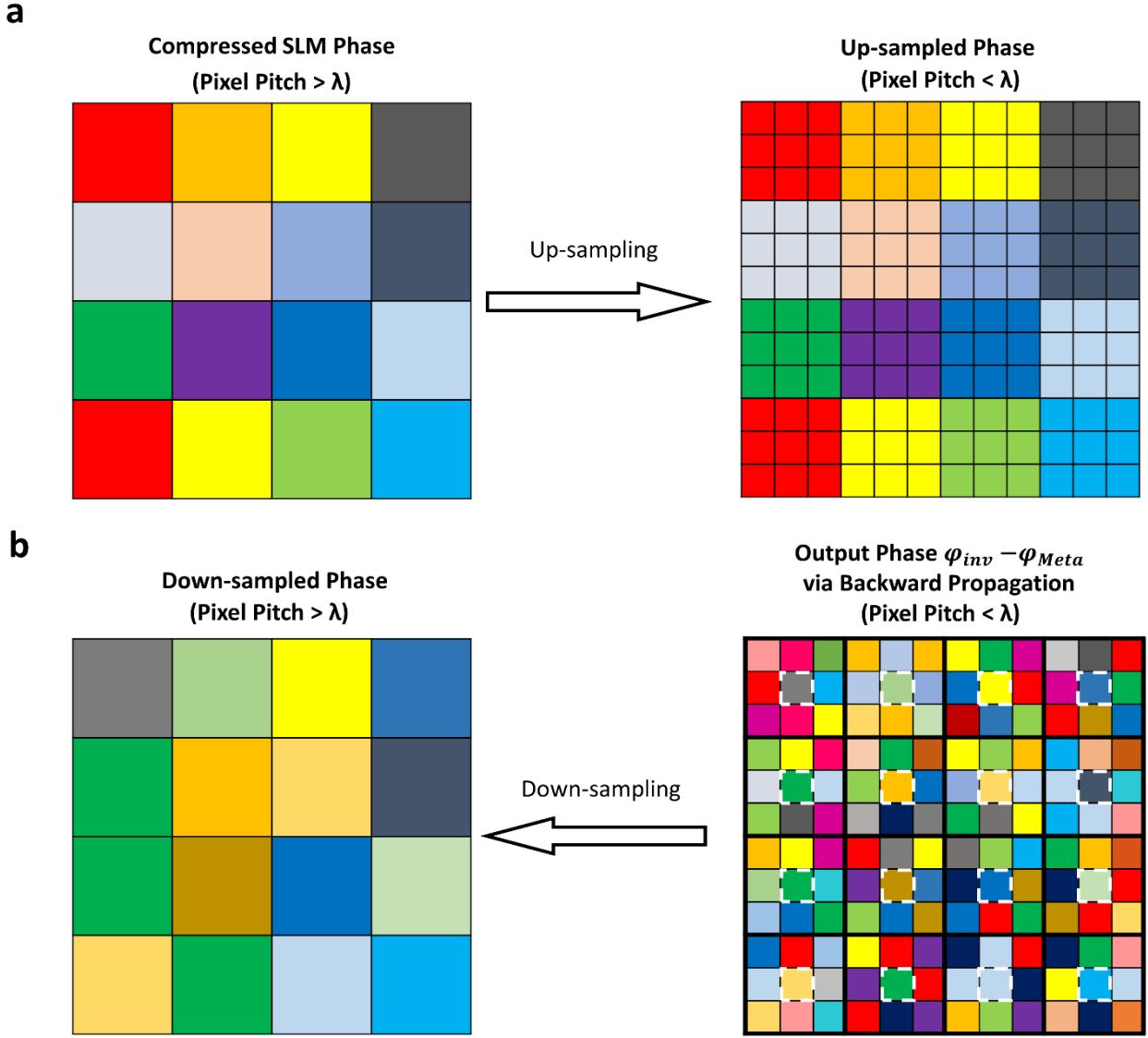

**Supplementary Figure 4. Sketch for up-sampling (a) and down-sampling (b) process of optimizing the SLM phase in our pixel-interpolation meta-holography.**

During the algorithm iteration process, a pixel matching algorithm is required to address the mismatch between the pixel pitch of the metasurface and that of the hologram. Its core function is to realize our proposed pixel-interpolation operation, thereby both phases can be re-generated smoothly in the forward and backward propagation of the optimization algorithm (see the sketch in Fig. 4a of main text).

First, we calculate the sampling ratio defined as $\eta = a*P_{SLM}/P_{META}$, where $a$ is the scaling factor of the imaging system, $P_{SLM}$ and $P_{META}$ represent the pixel sizes of the spatial light modulator (SLM) and the metasurfaces, respectively. To match the pixel pitch, the compressed SLM phase is up-sampled by dividing one original pixel into several subwavelength pixels, where each subwavelength pixel inherits the same phase from the original pixel (as sketched in Supplementary Fig. 4a). Thus, the interpolated SLM phase has the same sampling size with the metasurface phase so that the pixel-interpolation operation can be implemented correctly during the hologram design.

The down-sampling operation is carried out after backward propagation which yields the reversed phase with subwavelength pixel. The subwavelength-pixel phase is down-sampled by using the phase of only one pixel for each $N/M$, where $N$ and $M$ are the effective pixel numbers of the metasurfaces and the SLM along one direction, respectively. For instance, in our case of $N/M=3$, we take one every 3 pixels (see Supplementary Fig. 4b), thus creating a down-sampled phase of size $M \times M$.

### ■ Supplementary Section 6. Distortion pre-compensation of the target image

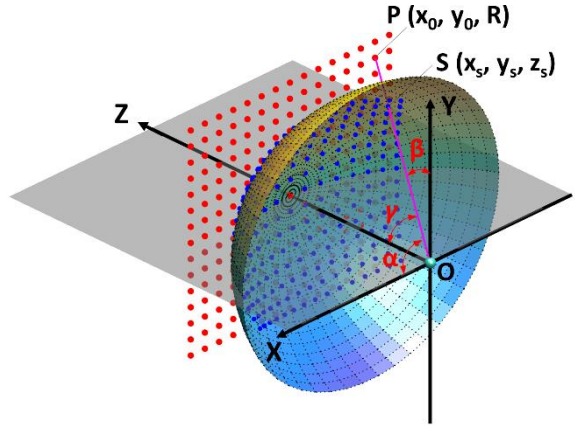

**Supplementary Figure 5. Sketch for the central projection from a plane to a sphere.** $\alpha$, $\beta$ and $\gamma$ are direction cosines between the line OP and three axes.

When the angular spectrum method is used to simulate the propagation of light, the diffracted pattern can be numerically calculated by a 2D-FFT transformation of the incident field multiplied by a quadratic phase. It means that the calculated pattern takes the coordinates of spatial frequency rather than spatial position. Under the Fresnel region, the spatial frequency can be approximated in terms of $k_x = k \cdot \alpha \approx k \cdot x/z$ and $k_y = k \cdot \beta \approx k \cdot y/z$, where the wave number $k=2\pi/\lambda$ ($\lambda$ is the operating wavelength), $x$, $y$ and $z$ are the spatial coordinates at the target plane. Only when the ratios $x/z$ and $y/z$ are small sufficiently, the Fresnel approximation is valid without the distortion. But, in our case, we simulate the diffracted patterns with up to 160° FOV, where the spatial frequencies are large so that $\alpha$ and $\beta$ must be calculated accurately. Thus, the target pattern in the spatial coordinates cannot be used as the ideal pattern for the large-FOV case. The correct method is to project the expected spatial pattern onto a sphere, where we can obtain a new pattern at the spatial frequency as the target of the Fresnel diffraction. This new pattern works as a pre-compensation of the distortion in the target plane, which holds the fundamental origins of distortion pre-compensation in this work.

Figure S5 sketches the central projection of a uniform dot array at the plane onto a sphere with a radius of $R$. For the sake of easy demonstration, the plane located at z is assumed to be tangential to the sphere with its origin of $O$. Following the central projection, we get the following geometric relationship

$$x = \frac{\alpha z}{\sqrt{1 - \alpha^2 - \beta^2}}, y = \frac{\beta z}{\sqrt{1 - \alpha^2 - \beta^2}}, \quad (S14)$$

where $\alpha$ and $\beta$ are direction cosines between the line OP and the x-axis and y-axis. By using Eq. (S14), we obtain

$$\frac{\partial x}{\partial \alpha} = z\left[\gamma^{-1} \cdot \frac{\partial}{\partial \alpha}(\alpha) + \alpha \cdot \frac{\partial}{\partial \alpha}(\gamma^{-1})\right] = z\left[\gamma^{-1} + \alpha \cdot (-\gamma^{-2})\frac{\partial \gamma}{\partial \alpha}\right] = z\gamma^{-3}(\gamma^2 + \alpha^2), \quad (S15a)$$

$$\frac{\partial x}{\partial \beta} = \alpha z \frac{\partial}{\partial \beta}(\gamma^{-1}) = \alpha z \cdot (-\gamma^{-2})\frac{\partial \gamma}{\partial \beta} = \alpha\beta z\gamma^{-3}, \quad (S15b)$$

$$\frac{\partial y}{\partial \alpha} = \beta z \frac{\partial}{\partial \alpha}(\gamma^{-1}) = \beta z \cdot (-\gamma^{-2})\frac{\partial \gamma}{\partial \alpha} = \beta z \cdot (-\gamma^{-2}) \cdot (-\alpha\gamma^{-1}) = \alpha\beta z\gamma^{-3}, \quad (S15c)$$

$$\frac{\partial y}{\partial \beta} = z\left[\gamma^{-1} \cdot \frac{\partial}{\partial \beta}(\beta) + \beta \cdot \frac{\partial}{\partial \beta}(\gamma^{-1})\right] = z\left[\gamma^{-1} + \beta \cdot (-\gamma^{-2})\frac{\partial \gamma}{\partial \beta}\right] = z\gamma^{-3}(\gamma^2 + \beta^2), \quad (S15d)$$

where $\gamma = (1 - \alpha^2 - \beta^2)^{1/2}$ is the direction cosines between the line OP along the z-axis. By employing the Jacobian determinant, we derive

$$J = \frac{\partial x}{\partial \alpha}\frac{\partial y}{\partial \beta} - \frac{\partial x}{\partial \beta}\frac{\partial y}{\partial \alpha} = [z\gamma^{-3}(1 - \beta^2)][z\gamma^{-3}(1 - \alpha^2)] - [\alpha\beta z\gamma^{-3}][\alpha\beta z\gamma^{-3}] = \frac{z^2}{\gamma^4}, \quad (S16)$$

$$dxdy = \frac{z^2}{(1 - \alpha^2 - \beta^2)^2}d\alpha d\beta = \frac{z^2}{\gamma^4}d\alpha d\beta \quad (S17)$$

Equation (S14) denotes the relationship between the spatial position of P and the spatial-frequency coordinates. One can find that there is a scaling factor of $\gamma$ from the spatial position ($x_0$, $y_0$) to the spatial frequency (α, β). The scaling factor ($\frac{z^2}{\gamma^4}$) in Eq. S17 results in the distorted shape of pattern at the spatial-frequency domain. By deriving the coordinate mapping relationship, we can achieve complete distortion correction.

■ **Supplementary Section 7. Experimental setup for FOV comparison**

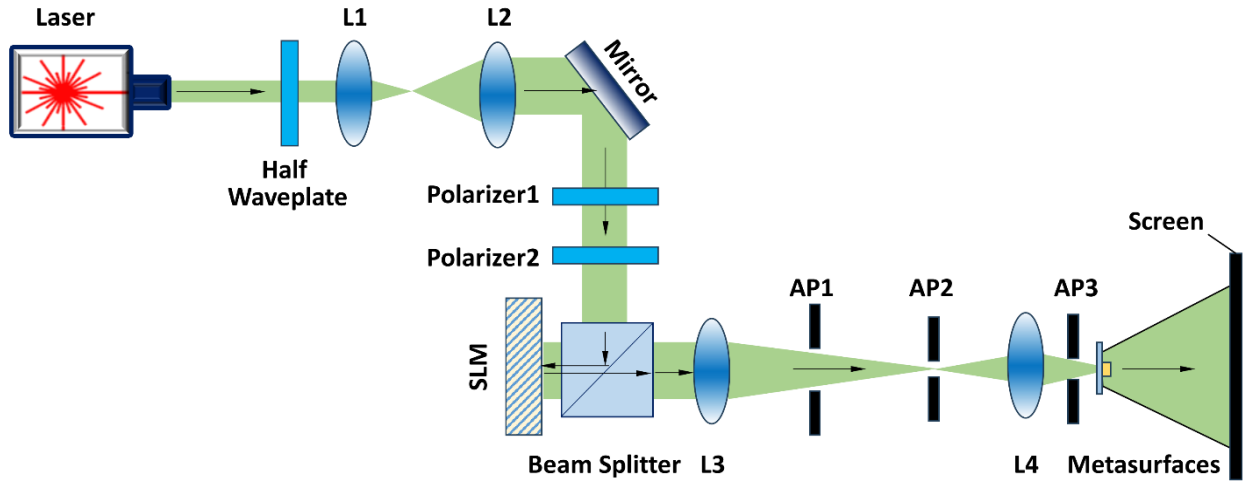

**Supplementary Figure 6. Experimental setup for characterizing the FOV.** AP: aperture, L: lens.

To demonstrate the metasurface's capability for expanding the FOV and its phase modulation performance, we build the experimental setup as sketched in Supplementary Fig. 6. In the optical path, a 561 nm laser beam (which is a linear polarized mode) passes through a half-wave plate to adjust the incident light's polarization state. The beam then traverses a 4-*f* beam expander system (comprising lenses $L_1$ and $L_2$) and two polarizers. The first polarizer is to adjust the power of the beam, while the second polarizer is to guarantee the proper linear polarized state for the SLM. The beam subsequently illuminates a reflective spatial light modulator (SLM) for phase modulation by a beam splitter (BS). After SLM phase modulation, the beam passes through the telescopic system (composed of lenses $L_3$ and $L_4$), where an aperture filters stray light. Optical beam modulated by SLM is compressed with a scaling factor of 5 (from 3.74 μm to 0.748 μm) to realize the pixel match between compressed SLM phase and the metasurface phase. Two apertures are added between lenses L3 and L4. The square aperture AP1 filters out areas where the spot exceeds the modulation range. Aperture AP2 performs low-pass filtering at the focal plane to eliminate stray light. An additional aperture (AP3) is placed between L4 and the metasurface to remove background light with large tilt angles. Light with compressed SLM phase illuminates on the metasurfaces for FOV expansion. The reconstructed images with large FOV are projected onto the display screen, thereby realizing large-FOV holography. The imaging distance between the metasurfaces and the screen is measured as z=8.39 mm, while the imaging size (see Fig. 5a in the main text) is 91.16×88.34 mm. Therefore, for metasurface holography, the FOV is

$$\text{FOV} = 2 \cdot \arctan\left(\frac{d}{2z}\right) = 2 \cdot \arctan\left(\frac{91.16}{2*8.39}\right) = 159.1°. \tag{S18}$$

SLM-only holographic characterization is also realized by the setup in Supplementary Fig. 6. Due to the phase profile difference, AP1-AP3 are enlarged. Considering the refractive index of the metasurface substrate, 0.25mm equivalent distance is added the diffraction distance. For the SLM-only case, the FOV (see Fig. 5b in the main text) is calculated as

$$\text{FOV} = 2 \cdot \arctan\left(\frac{3.4}{2*(8.39+0.25)}\right) = 22.3°. \tag{S19}$$

■ **References**